\documentclass[sigconf]{acmart}
\AtBeginDocument{%
  \providecommand\BibTeX{{%
    \normalfont B\kern-0.5em{\scshape i\kern-0.25em b}\kern-0.8em\TeX}}}

\copyrightyear{2024}
\acmYear{2024}
\setcopyright{rightsretained}
\acmConference[SIGIR '24]{Proceedings of the 47th International ACM SIGIR Conference on Research and Development in Information Retrieval}{July 14--18, 2024}{Washington, DC, USA}
\acmBooktitle{Proceedings of the 47th International ACM SIGIR Conference on Research and Development in Information Retrieval (SIGIR '24), July 14--18, 2024, Washington, DC, USA}
\acmDOI{10.1145/3626772.3657792}
\acmISBN{979-8-4007-0431-4/24/07}

\usepackage{hyperref}
\usepackage{algorithm}
\usepackage{algorithmic}

\usepackage{pifont}
\usepackage{newunicodechar}
\newunicodechar{✓}{\ding{51}}
\newunicodechar{✗}{\ding{55}}

\begin{document}

\title{Drop your Decoder: Pre-training with Bag-of-Word Prediction for Dense Passage Retrieval.}

% \author{
%     Guangyuan Ma\textsuperscript{\rm 1,2}, 
%     Xing Wu\textsuperscript{\rm 1,2}, 
%     Zijia Lin\textsuperscript{\rm 3},
%     Songlin Hu\textsuperscript{\rm 1,2}
% }

% \affiliation{
% \institution{
%     \textsuperscript{\rm 1}Institute of Information Engineering, Chinese Academy of Sciences, Beijing, China\\
%     \textsuperscript{\rm 2}School of Cyber Security, University of Chinese Academy of Sciences, Beijing, China\\
%     \textsuperscript{\rm 3}Kuaishou Technology, Beijing, China\\
%     \{maguangyuan,wuxing,husonglin\}@iie.ac.cn, linzijia07@tsinghua.org.cn
% }
% \country{}
% }

\author{Guangyuan Ma}
\email{maguangyuan@iie.ac.cn}
\orcid{0000-0001-6916-9611}
\affiliation{
  \institution{Institute of Information Engineering, \\Chinese Academy of Sciences}
  \department{School of Cyber Security, University \\of Chinese Academy of Sciences}
  \city{Beijing}
  % \state{Beijing}
  \country{China}
  \postcode{100000}
}

\author{Xing Wu}
\email{wuxing@iie.ac.cn}
\orcid{0009-0004-3796-3705}
\affiliation{
  \institution{Institute of Information Engineering, \\Chinese Academy of Sciences}
  \department{School of Cyber Security, University \\of Chinese Academy of Sciences}
  \city{Beijing}
  % \state{Beijing}
  \country{China}
  \postcode{100000}
}

\author{Zijia Lin}
\email{linzijia07@tsinghua.org.cn}
\orcid{0000-0002-1390-7424}
\affiliation{
  \institution{Tsinghua University}
  \city{Beijing}
  % \state{Beijing}
  \country{China}
  \postcode{100000}
}

\author{Songlin Hu}
\authornote{Corresponding Author}
\email{husonglin@iie.ac.cn}
\orcid{0000-0002-7170-3809}
\affiliation{
  \institution{Institute of Information Engineering, \\Chinese Academy of Sciences}
  \department{School of Cyber Security, University \\of Chinese Academy of Sciences}
  \city{Beijing}
  % \state{Beijing}
  \country{China}
  \postcode{100000}
}

\renewcommand{\shortauthors}{Ma, et al.}

\begin{abstract}
Masked auto-encoder pre-training has emerged as a prevalent technique for initializing and enhancing dense retrieval systems. It generally utilizes additional Transformer decoder blocks to provide sustainable supervision signals and compress contextual information into dense representations. However, the underlying reasons for the effectiveness of such a pre-training technique remain unclear. The usage of additional Transformer-based decoders also incurs significant computational costs. In this study, we aim to shed light on this issue by revealing that masked auto-encoder (MAE) pre-training with enhanced decoding significantly improves the term coverage of input tokens in dense representations, compared to vanilla BERT checkpoints. Building upon this observation, we propose a modification to the traditional MAE by replacing the decoder of a masked auto-encoder with a completely simplified Bag-of-Word prediction task. This modification enables the efficient compression of lexical signals into dense representations through unsupervised pre-training. Remarkably, our proposed method achieves state-of-the-art retrieval performance on several large-scale retrieval benchmarks without requiring any additional parameters, which provides a 67\% training speed-up compared to standard masked auto-encoder pre-training with enhanced decoding. 
% \footnote{Our code will be available at \url{https://github.com/ma787639046/bowdpr}.}
\end{abstract}






\maketitle

\section{Introduction}
% Passage retrieval aims to search relevant passages from large text collections based on the similarity search with given queries, which has broad applications in real-world scenarios such as search engine \cite{lu2022ernie_search}, open domain question-answering \cite{karpukhin2020dpr}. 
Dense passage retrieval utilizes pre-trained language models (PLMs), e.g., BERT \cite{devlin2019bert}, to encode the lexical information as representations in latent spaces and retrieve relevant passages of a given query based on the similarity search \cite{Stephen2016MIPS, Yury2020HNSW}. Pre-training with masked auto-encoder (MAE) has recently become a hot topic to boost the initialization ability of PLMs for downstream retrieval tasks. Such a technique appends carefully designed Transformers-based decoder blocks to the PLM encoder, forming encoder-decoder architecture, and bringing supervision signals to the representations. 

\begin{figure}[tbp!]
\centering
\includegraphics[width=\linewidth]{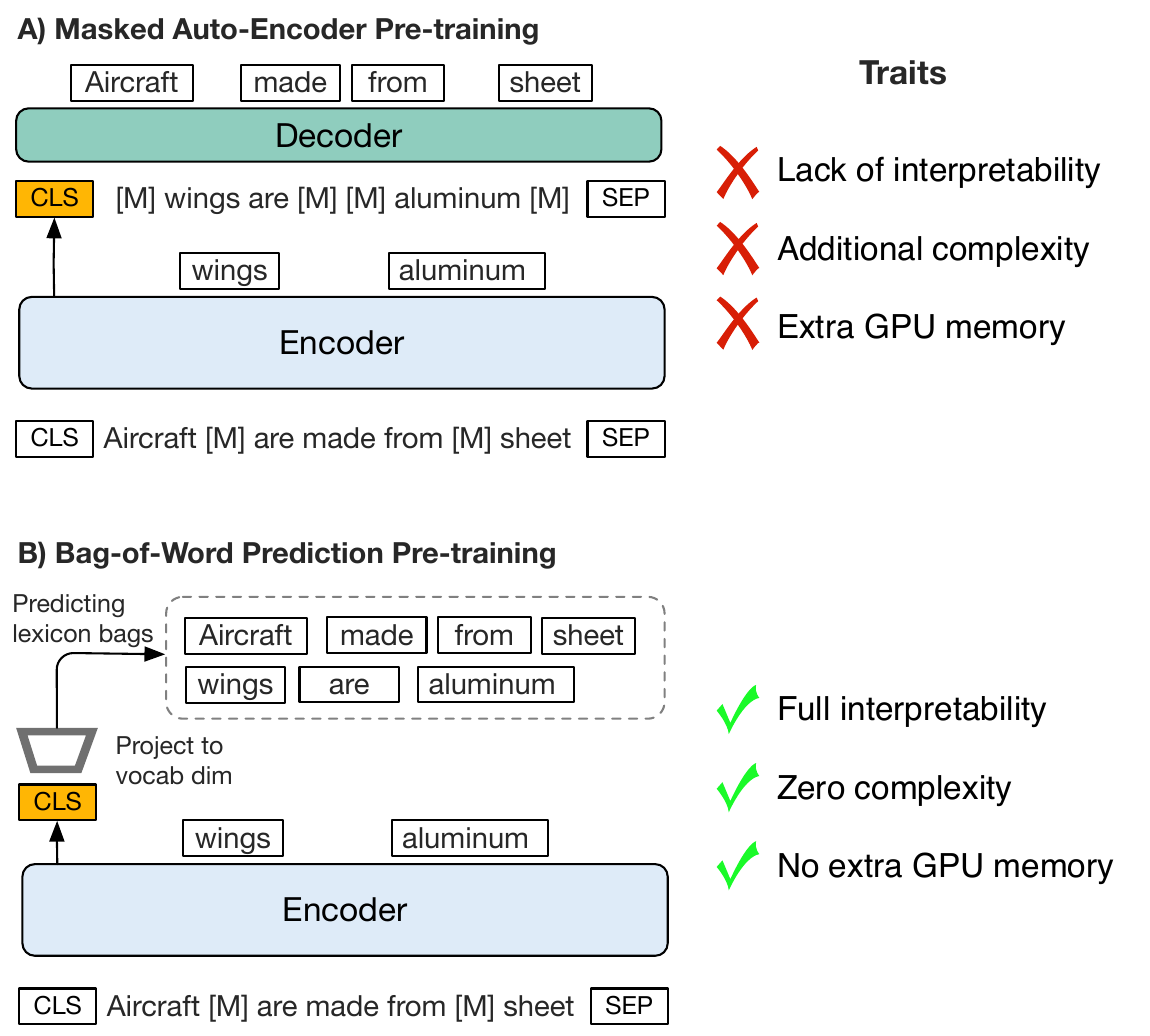}
\caption{
Comparison of Masked Auto-Encoder Pre-training and Bag-of-Word Prediction Pre-training.
}
\label{bow_fig1}
\end{figure}

Recent studies have explored several pre-training strategies, such as pre-training with the weakened decoder \cite{lu2021less}, extreme mask ratio \cite{liu2022retromae}, replace token detection \cite{wang2022simlm}, context reconstruction \cite{wu2023contextual}, and multi-task decoding \cite{zhou2022master}. As is shown in Figure \ref{bow_fig1} A), these MAE-style pre-trainings commonly preserve several traits: 1) \textbf{Asymmetric structure}: The encoder is a full-sized Transformers-based encoder for generating discriminative representations based on input sentences. While the decoder is typically a shallow Transformers-based block with a few or just one layer. 2) \textbf{Bottlenecked connection}. The connection between the encoder and the decoder is typically a single representation vector. 3) \textbf{Reconstruction signals}. Carefully designed decoding tasks are applied to the decoder side, which decodes context information through auto-encoding \cite{gao2021condenser, liu2022retromae}, auto-regression \cite{lu2021less}, enhanced decoding \cite{liu2022retromae} or other curated decoding tasks \cite{wang2022simlm, zhou2022master, wu2022query-as-context}. By optimizing a weak decoder, the lexical information is compressed into the sentence representations, thus providing good initialization ability for retrieval tasks. Notably, enhanced decoding from \cite{liu2022retromae} is proven effective for dense retrieval pre-training, which utilizes sentence representations as query streams and masked context embeddings as content streams. 

However, we argue that existing MAE-style pre-training paradigms have several drawbacks, which may hinder the development of dense retrieval pre-training: 1) \textbf{Lack of interpretability}. Existing studies focus on creating carefully designed tasks, but still neglect to reveal the underlying reason for why such a pre-training method is effective. 2) \textbf{Additional complexity}. MAE-style pre-training involves optimization with millions of document collections. The usage of additional Transformers-based decoders brings considerable GPU memory usage and additional $O(n^{2})$ computational complexity. Some studies also need additional data preprocessing \cite{wu2022query-as-context, zhou2022master} or extensive data masking \cite{liu2022retromae, wu2023contextual}.

In this study, we aim to solve the above issues and develop a more computationally efficient and more interpretable pre-training technique. The semantic meaning of a dense representation is typically condensed in latent spaces and hard to directly interpret. In our preliminary studies, we project the low-dimensional representation to the vocabulary space with the language model heads and compare the changes in MAE-style pre-training. Notably, we discover that MAE-style pre-training with enhanced decoding \cite{liu2022retromae} significantly enhances the term coverage of original input tokens, which implies that the stronger retrieval ability may come from the compression of input lexicon terms into dense representations. 

Building upon this observation, we propose a new pre-training schema, the Bag-of-Word prediction task, for directly compressing the lexicon terms into dense representations. As is shown in Figure \ref{bow_fig1} B), it first projects dense representations to the vocabulary space, then directly predicts the input token bags with multi-label cross-entropy loss. As a result, the lexical information of input tokens is directly compressed into the dense representations, which will give superior initialization ability to the PLMs for downstream retrieval tasks. This method is a straightforward replacement for existing MAE-style pre-training paradigms. There are several benefits to utilizing a Bag-of-Word prediction task, instead of previous MAE-style pre-training. 

\noindent $\bullet$ \textbf{Zero computational complexity.} This method requires \textbf{NO} extra Transformers-based decoder layers, GPU memory consumption, data preprocessing, or extensive data masking. It is easy to implement and greatly reduces the additional computational cost of the pre-training. 

\noindent $\bullet$ \textbf{High interpretability.} Compared to complex modifications to the decoder tasks \cite{liu2022retromae, wang2022simlm, wu2023contextual, zhou2022master} of MAE-style pre-training, our method removes the need for adding decoders and directly optimizes a multi-label prediction task of input lexicons. 

To verify the effectiveness of our method, we conduct experiments on several large-scale web search and open-domain QA datasets, including MS-MARCO passage ranking \cite{tri2016msmarco}, Natural Questions (NQ) \cite{tom2019nq}, TriviaQA \cite{Mandar2017TriviaQA} and heterogeneous zero-shot BEIR benchmarks \cite{thakur2021beir}. Our proposed method achieves state-of-the-art retrieval performance and provides a 67\% training speed-up compared to masked auto-encoder with enhanced decoding \cite{liu2022retromae}.

In summary, our contributions are as follows.

\noindent $\bullet$ We first reveal that masked auto-encoder (MAE) pre-training with enhanced decoding significantly improves the term coverage of input tokens in dense representations.

\noindent $\bullet$ Building upon this observation, we propose a direct Bag-of-Word prediction task for compressing the information of input lexicon terms into dense representations.

\noindent $\bullet$ Our method achieves considerable training speed-up without additional computational complexity, and meanwhile keeping full interpretability as well as state-of-the-art retrieval performances on multiple retrieval benchmarks.

\section{Related Works}
\paragraph{\textbf{Passage Retrieval.}}
Passage retrieval aims to find relevant information from large passage collections based on a given search query. Traditional techniques for ad-hoc retrieval, such as TF-IDF and BM25 \cite{robertson2009probabilistic}, perform relevant searches based on lexicon terms and inverted document frequencies. Such methods have efficient retrieval abilities, but they suffer from vocabulary mismatch problems \cite{karpukhin2020dpr}. Although there are continuous techniques like query expansions \cite{Rodrigo2019doc2query, wang2023Query2doc, jagerman2023QueryExpension} and query rewrite \cite{Sreenivas2011query_rewrite} to alleviate such mismatch issues, additional preprocess of queries or passage collections make the retrieval pipelines much more complicated and hard to maintain. 

Dense retrieval \cite{karpukhin2020dpr, qu-etal-2021-rocketqa} has become popular in mainstream retrieval applications recently, because it maps the textual information into learnable representations and naturally alleviates the lexicon mismatch problems by learning the similarity traits of queries and passages through end-to-end contrastive learning. Dense retrieval encodes queries and passage collections as condensed low-dimension sentence representations with pre-trained language models (PLMs), and performs retrieval based on the similarity of representations with efficient search algorithms like maximum inner-product search (MIPS) \cite{Stephen2016MIPS}. Notably, there are continuous efforts \cite{Thibault2021splade,Thibault2021splade_v2,shen2023lexmae,yu2024corpus_specific_vocabularies} to adapt the PLMs to sparse retrieval. There are also extended studies about hybrid retrievals \cite{liu2023retromaev2, wu2023cotmaev2}, which combines both dense and sparse retrievals to bring superior retrieval performances. However, our work primarily focuses on dense retrieval pre-training and is orthogonal to these works.

\begin{table}[!t]
\centering
\resizebox{\linewidth}{!}{
    \begin{tabular}{l|c|cccc}
    \toprule  
    ~ & \textbf{Eval} & \multicolumn{4}{c}{\textbf{MS-MARCO}} \\
    \textbf{Model} & \textbf{MLM Loss} $\downarrow$ & MRR@10 $\uparrow$ & R@1 $\uparrow$ & R@50 $\uparrow$ & R@1k $\uparrow$ \\
    \midrule
    BERT & 2.9389  & 36.05  & 23.55  & 83.48  & 96.93   \\ 
    Further Pre-train & \textbf{1.8863}  & 37.33 $\dagger$  & 24.90 $\dagger$  & 84.81 $\dagger$  & 97.66 $\dagger$ \\
    Auto-Encoding & 1.9499  & 37.72  & 25.06  & 86.16  & 98.02   \\
    Auto-Regression & 1.9765  & 37.06  & 24.71  & 84.76  & 97.56   \\
    Enhanced Decoding & 2.0027  & \textbf{38.12} \textsuperscript{*}  & \textbf{25.19}  & \textbf{86.49} \textsuperscript{*}  & \textbf{98.22} \textsuperscript{*}  \\
    \bottomrule
    \end{tabular}
}
\caption{
MLM Loss and retrieval performances of MAE-style pre-training models on the MS-MARCO passage ranking dataset. Better results are marked as bold. $\dagger$ means significant improvement over ``\texttt{BERT}'' baselines. \textsuperscript{*} means significant improvement over ``\texttt{Further Pre-train}'' settings. ($p \leq 0.05$)
}
\label{table_mae_eval_loss_msmarco}
\end{table}

\paragraph{\textbf{Pre-training for Dense Retrieval.}}
There are continuous efforts \cite{gao2021condenser, gao2022unsupervised, ren-etal-2021-rocketqav2, lu2022ernie_search} to improve the retrieval quality of sentence representations with specifically designed pre-training techniques, where masked auto-encoder (MAE) \cite{gao2021condenser, lu2021less, gao2022unsupervised, liu2022retromae, wu2023contextual, wang2022simlm, zhou2022master} has gained popularity to boost the initialization of PLMs for downstream retrieval tasks in recent years. Masked auto-encoder (MAE) pre-training seeks to compress the lexicon information of contextual texts into sentence representations through various auxiliary pre-training tasks and tailored bottlenecked architectures. It usually appends a shallow decoder to the PLM encoder and connects the encoder-decoder with merely a sentence representation. Through the process of MAE-style pre-training, the information of input texts is first condensed as sentence representations and then decoded in the appended decoder with auto-encoding \cite{gao2021condenser, gao2022unsupervised, liu2022retromae, wu2023contextual, zhou2022master}, auto-regression \cite{lu2021less} or enhanced decoding \cite{liu2022retromae}. Such an information encoding-decoding process brings sustainable supervision signals to the sentence representations and thus gives better initialization to the PLMs for downstream retrievals.

Recent studies focus on designing various decoding tasks for MAE-based pre-trainings. \cite{lu2021less} proposes to pre-train an encoder with merely a single-layer auto-regressive decoder \cite{radford2019gpt2} and apply attention window restriction to enforce textual information compression to the representations. \cite{gao2021condenser, gao2022unsupervised} divides the MAE into early-late-head layers and proposes to optimize with condensed representations and structural readiness. Furthermore, \cite{liu2022retromae, liu2023retromaev2} proposes pre-training with an auto-encoding decoder and an extreme mask ratio. They further propose an enhanced decoding paradigm \cite{zhilin2019xlnet} for improving token usage ability in MAE-style pre-training. It applies sentence representations with positional information as query streams and performs cross-attention with masked context embeddings as content streams in the decoder. In our studies below, we observe significantly increased term coverage of input lexicon tokens by projecting the pre-trained sentence representations of enhanced decoding to the vocabulary space. This implies that the stronger retrieval ability may be attributed to the compression of lexicon terms into dense representations.

\begin{table}[!t]
\centering
\resizebox{\linewidth}{!}{
    \begin{tabular}{l|c|cc|cc}
    \toprule  
    ~ & \textbf{Eval} & \multicolumn{2}{c|}{\textbf{NQ}} & \multicolumn{2}{c}{\textbf{TriviaQA}} \\
    \textbf{Model} & \textbf{MLM Loss} $\downarrow$ & R@20 $\uparrow$ & R@100 $\uparrow$ & R@20 $\uparrow$ & R@100 $\uparrow$ \\
    \midrule
    BERT & 2.9498  & 83.13  & 88.78  & 82.21  & 86.83   \\
    Further Pre-train & \textbf{1.7682}  & 83.99 $\dagger$ & 89.67 $\dagger$ & 83.79 $\dagger$ & 87.62 $\dagger$  \\ 
    Auto-Encoding & 1.8320  & \textbf{84.52} \textsuperscript{*} & 90.00  & 84.02  & 87.61   \\ 
    Auto-Regression & 1.8609  & 84.40 \textsuperscript{*} & 90.06  & \textbf{84.63} \textsuperscript{*} & 87.93   \\ 
    Enhanced Decoding & 1.8884  & 84.46 \textsuperscript{*} & \textbf{90.30} \textsuperscript{*} & 84.51 \textsuperscript{*} & \textbf{88.08}  \\ 
    \bottomrule
    \end{tabular}
}
\caption{
MLM Loss and retrieval performances of MAE-style pre-training models on Natural Questions (NQ) \cite{tom2019nq}, TriviaQA \cite{Mandar2017TriviaQA} datasets.
}
\label{table_mae_eval_loss_dpr}
\end{table}

\paragraph{\textbf{Drawbacks of Existing MAE Pre-training.}}
Although the bloom of MAE-style pre-training gives better retrieval performances to dense systems, several obvious drawbacks hinder the further development of such pre-training techniques. Firstly, existing MAE pre-trainings focus on creating carefully designed auxiliary pre-training tasks, but it's hard to interpret the underlying reason why such an auxiliary pre-training task is effective. Furthermore, MAE-style pre-training mandatory needs the additional Transformers-based decoders, which brings considerable GPU memory cost and additional $O(n^{2})$ computational complexity. Complex data preprocessing \cite{wu2022query-as-context, zhou2022master} or extensive data masking \cite{liu2022retromae, wu2023contextual} are also needed in several implementations. In contrast, our method breaks the constraints of using any additional decoders, preprocessing, or masking, by directly optimizing a Bag-of-Word decoding task. Such a method achieves state-of-the-art retrieval performances on various datasets. It is highly interpretable and has nearly zero additional computational complexity in the whole pre-training process. Note that some works also involve term prediction tasks \cite{shengyao2021tilde, herve2023benchmarking_middle}. However, \cite{shengyao2021tilde} utilizes the query-document likelihoods in the fine-tuning stage, while our work utilizes a simpler Bag-of-Word prediction task without query-document interaction. Our study also focuses on the interpretability of MAE pre-training for dense retrieval and optimization in unsupervised pre-training. \cite{herve2023benchmarking_middle} makes performance comparisons with recent middle training methods and augments with the prediction of L2 normalized frequencies of the input tokens. In comparison, our work reveals the underlying mechanism of performance gains for MAE pre-training and proposes a simple Bag-of-Word prediction instead of frequency prediction.

\section{Method}
In this section, we first introduce the preliminary of dense retrieval and the drawbacks of existing masked auto-encoder pre-training. Then we analyze the interpretability of MAE pre-training in detail. Next, we introduce a new Bag-of-Word pre-training schema for efficient dense retrieval pre-training.

\subsection{Preliminary of Dense Retrieval Pre-training}
Dense retrieval utilizes an encoder $Enc$, e.g., BERT \cite{devlin2019bert}, to generate the sentence representations $v_q$ and $v_p$ of a query $q$ and a passage $p$, respectively. Typically, the representations are derived from the last hidden states at the \texttt{[CLS]} token position, i.e., $h_{last}^{CLS}$. Then the similarity $Sim(q, p)$ of $v_q$ and $v_p$ is calculated with dot product or cosine similarity.

\begin{equation}
\label{equation_query_passage_similarity}
    Sim(q, p) = Enc(q) \cdot Enc(p) = v_q \cdot v_p
\end{equation}

Masked auto-encoder (MAE) pre-training is an effective schema to bring sustainable supervision signals to sentence representations of the encoder and boost the retrieval performance for subsequent fine-tuning. Given a tokenized text $t \in T$ from training corpus collections, MAE pre-training first selects a proportion of tokens, with a pre-defined mask ratio of $r_{Enc}$, and replaces each with a mask token \texttt{[m]}.

\begin{equation}
    mask(t) = \{[CLS], t_1, t_2, [m], t_4, ..., t_n, [SEP]\}
\end{equation}

Then the masked texts $mask(t)$ are forwarded through the encoder $Enc$ to optimize a Masked Language Model (MLM) task loss $\mathcal{L}_{enc}$ with Cross-Entropy (CE) loss.

\begin{equation}
\begin{aligned}
    \mathcal{L}_{Enc} &= CE(Enc(mask({t})), t_i), i \in M \\
    &= -\sum_{t\in T}\sum_{i \in {M}} \log p(t_i|Enc(mask({t})))
\end{aligned}
\end{equation}
where $M$ denotes the set of masked tokens.

Decoder designs are key innovations of MAE-style pre-training. There are mainly three categories of decoder designs, including auto-encoding \cite{gao2021condenser, liu2022retromae}, auto-regression \cite{lu2021less}, and enhanced decoding \cite{liu2022retromae}.

\begin{algorithm}[!t]
    %\textsl{}\setstretch{1.8}
    \renewcommand{\algorithmicrequire}{\textbf{Input:}}
    \renewcommand{\algorithmicensure}{\textbf{Output:}}
    \renewcommand{\algorithmiccomment}[1]{\hfill $\triangleright$ #1}
    \caption{Input Token Coverage Ratio}
    \label{algorithm_input_token_coverage}
    \begin{algorithmic}[1]
        \REQUIRE Projected Representation $h_{proj}$, Tokenized Inputs $T$
        \STATE $T_{topk} \leftarrow$ Top-$k$ dominating tokens from $h_{proj}$ 
        \STATE $\mathrm{r}_{dominate} \leftarrow$ Numbers of $(T_{topk} \cap T)$ / Numbers of $T$
        \ENSURE Input Token Coverage Ratio $\mathrm{r}_{dominate}$
    \end{algorithmic}  
\end{algorithm}

\begin{algorithm}[!t]
    %\textsl{}\setstretch{1.8}
    \renewcommand{\algorithmicrequire}{\textbf{Input:}}
    \renewcommand{\algorithmicensure}{\textbf{Output:}}
    \renewcommand{\algorithmiccomment}[1]{\hfill $\triangleright$ #1}
    \caption{Bag-of-word Prediction}
    \label{algorithm_bag_of_word}
    \begin{algorithmic}[1]
        \REQUIRE Tokenized Inputs $T$
        \STATE $h_{last}^{CLS} \leftarrow Enc(mask(T))$ \COMMENT{Encoder Forward}
        \STATE $h_{proj} \leftarrow h_{last}^{CLS} \cdot E^T $ \COMMENT{Projection}
        \STATE $\mathcal{L}_{Dec} \leftarrow CrossEntropy(h_{proj}, T)$ \COMMENT{Equation \ref{euqation_bow}}
        \ENSURE Bag-of-word Prediction Loss $\mathcal{L}_{Dec}$
    \end{algorithmic}  
\end{algorithm}

\paragraph{\textbf{Auto-Encoding Decoder}}
Given the text $t_{Dec}$ as a decoder input, the auto-encoding decoder first masks it with a large mask ratio $r_{Dec} (r_{Dec} > r_{Enc})$, and replaces the first token with the encoder representations.

\begin{equation}
    mask(t_{Dec}) = \{h_{last}^{CLS}, t_1, t_2, [m], t_4, ..., t_n, [SEP]\}
\end{equation}

Then the decoder also applies the MLM loss as reconstruction signals and optimizes the encoder representation through the gradient flow of the bottleneck-connected encoder-decoder architecture.

\begin{equation}
    \mathcal{L}_{Dec} = CE(Dec(mask(t_{Dec})), t_i), i \in M
\end{equation}

\begin{figure}[t!]
\centering
\includegraphics[width=\linewidth]{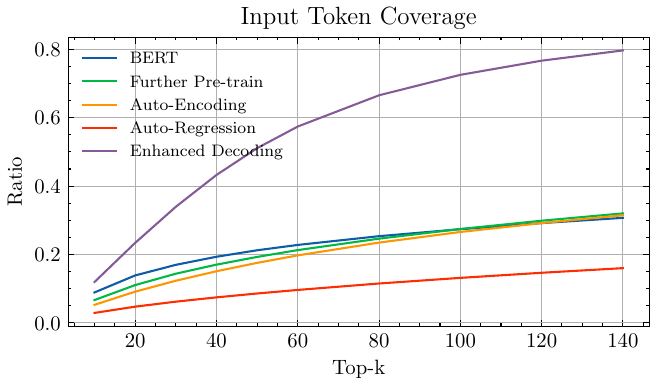}
\caption{
Compositions of Top-k tokens of dense representation.
}
\label{token_logits_01}
\end{figure}

\begin{figure*}[ht!]
\centering
\includegraphics[width=\linewidth]{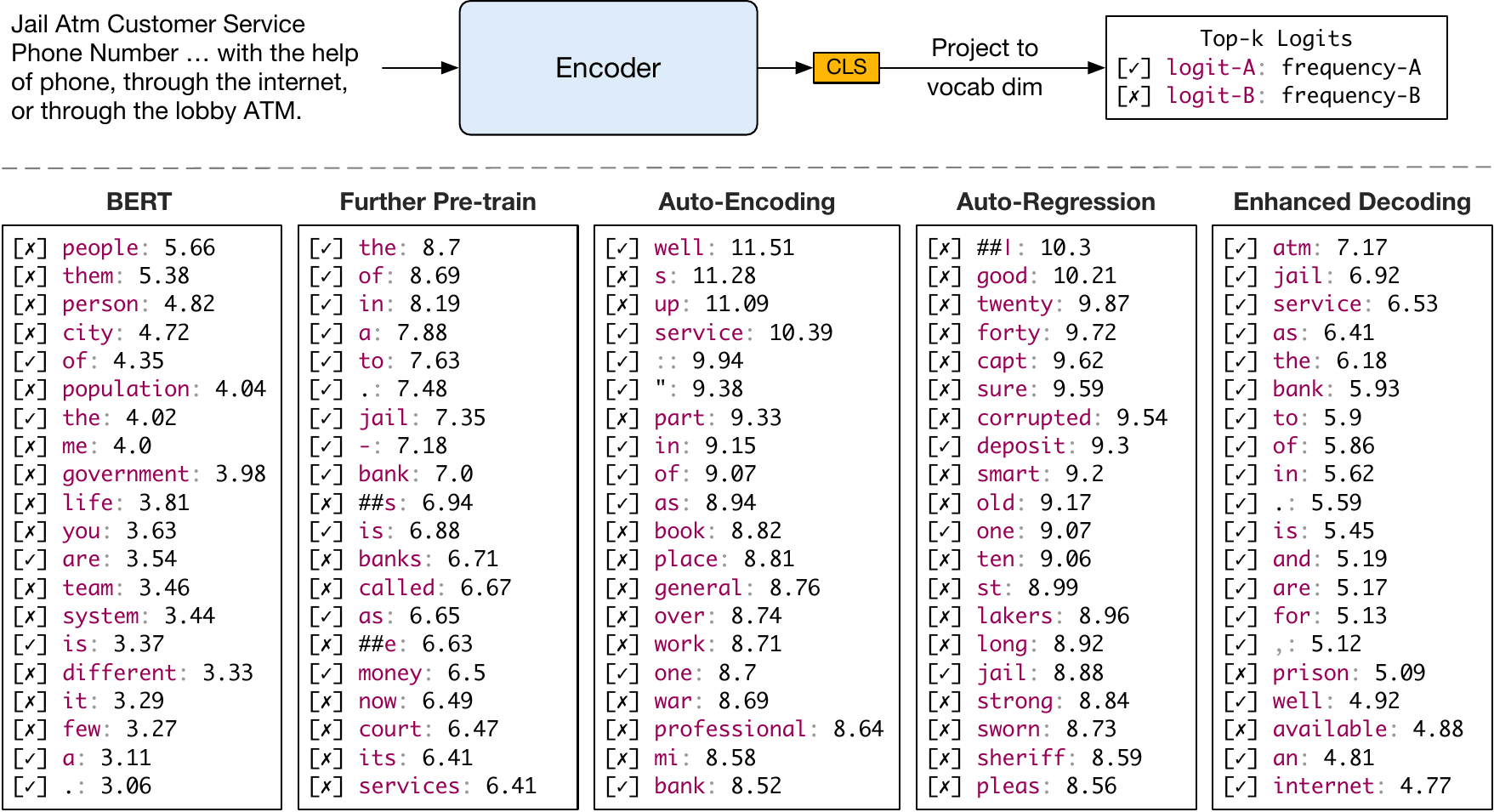}
\caption{
Examples of the compositions of Top-20 tokens of dense representation. We encode the input texts with various encoders and project them to the vocabulary space to interpret dominating tokens. ✓ means a token hits the input text, while ✗ means miss of the input text. 
}
\label{logits_view}
\end{figure*}

\paragraph{\textbf{Auto-Regression Decoder}}
Given the decoder input text $t_{Dec}$, it also replaces the first token with the encoder representations.

\begin{equation}
    t_{Dec} = \{h_{last}^{CLS}, t_1, t_2, t_3, t_4, ..., t_n, [SEP]\}
\end{equation}

Then Casual Language Model (CLM) loss is used as the generation signal for the decoder, implemented with the cross-entropy (CE) loss. The encoder representation will be optimized via the gradient flow of decoder generation from the decoder to the encoder.

\begin{equation}
\begin{aligned}
    \mathcal{L}_{Dec} &= CE(Dec(t_{Dec}[:i-1]), t_i), i \in (0, n] \\
    &= -\sum_{t_{i} \in {T}_{Dec}} \log p(t_{i}|Dec(t_{Dec}[:i-1]))
\end{aligned}
\end{equation}
where $t_{Dec}[:i-1]$ denotes all the tokens before $t_i$.

\paragraph{\textbf{Enhanced Decoding Decoder}}
Unlike self-attention schemas with auto-encoding or auto-regression, enhanced decoding \cite{liu2022retromae, zhilin2019xlnet} develops a cross-attention pattern with an extreme masked ratio for improving token usage. The decoder inputs $t_{Dec}$ are first forwarded through the embedding layer and converted to input embeddings $h_{Dec}^{Emb}$. Then enhanced decoding utilizes encoder representations $h_{last}^{CLS}$ with positional embeddings $p$ as the query stream $q$, and input embeddings $h_{Dec}^{Emb}$ as the content stream $k$ and $v$.

\begin{equation}
\begin{aligned}
    q = h_{last}^{CLS} + p \\
    k = v = h_{Dec}^{Emb}
\end{aligned}
\end{equation}

Afterward, a cross-attention is performed with the streams $q$, $k$, and $v$, then forwarded through feed-forward network (FFN) chunks with post-layer norm (LN).

\begin{equation}
\begin{aligned}
    A = softmax(q^T k / \sqrt{d} + M) \cdot v \\
    h_{Dec} = LN(FFN(A) + A)
\end{aligned}
\end{equation}

Specifically, a position-specific attention mask $M_{ij} \in [L, L]$ ($L$ is the sequence length) is applied to the cross-attention to increase the difficulty of the decoder optimization objective. This mask $M_{ij}$ is composed of a diagnose positional mask $i=j$ and a randomly sampled token mask $i, j \in sample([0...L])$. After projecting the decoder's hidden states $h_{Dec}$ with BERT language-model heads (i.e., its embedding matrix $E^T$), a cross-entropy loss is used with all corresponding input tokens as labels. 

\begin{equation}
\begin{aligned}
    \mathcal{L}_{Dec} = CE(h_{Dec} E^T, t_i), i \in [0, n]
\end{aligned}
\end{equation}

The above three decoder designs are popular schemas for masked auto-encoder (MAE) pre-training of dense retrieval. They provide rich supervision signals to the encoder representation and enable better initialization for downstream retrieval tasks. However, MAE pre-training lacks interpretability for its effectiveness and brings additional $O(n^2)$ complexity because of the mandatory need for extra Transformers-based decoder layers. These drawbacks hinder the further development of the existing MAE-style pre-training.

\subsection{Interpret MAE Pre-training}
\label{method_interprete}
\paragraph{\textbf{Adaptation Abilities and Retrieval Performances}}
The MLM loss on target corpora reflects the modeling abilities of various pre-trained checkpoints. To evaluate the domain adaptation abilities of the PLMs on the target corpus, we randomly sample 10,000 passages from MS-MARCO \cite{tri2016msmarco} and DPR-Wiki \cite{karpukhin2020dpr} collections and calculate the corresponding average MLM loss. 

Here we involve 5 models for performance comparison. 1) Original BERT checkpoint \cite{devlin2019bert}; 2) The BERT further pre-trained with MLM on the original training corpus, i.e., Wikipedia and Bookcorpus collections; 3) MAE Pre-training with Auto-Encoding Decoder; 4) MAE Pre-training with Auto-Regression Decoder; 5) MAE Pre-training with Enhanced Decoding Decoder. For fair comparisons, we mainly reproduce the pre-training settings in \cite{liu2022retromae} and directly take the enhanced decoding checkpoint from it. The MLM loss on evaluation sets of MS-MARCO, DPR-Wiki, and the corresponding retrieval performances are listed in Table \ref{table_mae_eval_loss_msmarco} and \ref{table_mae_eval_loss_dpr}.

Here we can observe some interesting phenomena. 1) Both further pre-training with pure MLM loss and MAE-style pre-training significantly lower the MLM loss on the target evaluation corpus, and improve the downstream retrieval performances. 2) 
Although MAE-style pre-training has higher MLM loss than the further pre-trained checkpoint, they tend to achieve higher retrieval performances. Thus we can conclude that pre-training for dense retrieval is effective for better domain adaptation abilities and retrieval performances, but such abilities are not always positively correlated to lower MLM losses on target corpora when using MAE-style pre-training. Thus, other critical traits should contribute to better retrieval abilities for MAE-style pre-training.

\begin{table*}[!ht]
\centering
% \resizebox{\linewidth}{!}{
\begin{tabular}{l|c|ccc|ccc|ccc}
\toprule
    ~ & \textbf{Additional} & \multicolumn{3}{c|}{\textbf{MS-MARCO}} & \multicolumn{3}{c|}{\textbf{Natural Question}} & \multicolumn{3}{c}{\textbf{Trivia QA}}  \\ 
    \textbf{Model} & \textbf{Decoder} & MRR@10 & R@50 & R@1k & R@5 & R@20 & R@100 & R@5 & R@20 & R@100  \\ 
    \midrule
    BM25 \cite{robertson2009probabilistic} & ~ & 18.7  & 59.2  & 85.7  & - & 59.1  & 73.7  & - & 66.9  & 76.7   \\ 
    SPLADE++ \cite{Thibault2022splade++} & ~ & 38.0  & - & 98.2  & - & - & - & - & - & -  \\ 
    \midrule
    DPR \cite{karpukhin2020dpr} & ~ & 31.7  & 58.0  & 85.7  & - & 74.4  & 85.3  & - & 79.3  & 84.9   \\ 
    ANCE \cite{xiong2020ance} & ~ & 33.0  & - & 95.9  & - & 81.9  & 87.5  & - & 80.3  & 85.3   \\ 
    SEED \cite{lu2021less} & $\checkmark$ & 33.9  & - & 96.1  & - & 83.1  & 88.7  & - & - & -  \\ 
    Condenser \cite{gao2021condenser} & $\checkmark$ & 36.6  & 85.2  & 97.4  & 73.5  & 83.2  & 88.4  & 74.3  & 81.9  & 86.2   \\ 
    RocketQA \cite{qu-etal-2021-rocketqa} & ~ & 37.0  & - & 97.9  & 74.0  & 82.7  & 88.5  & - & - & -  \\ 
    coCondenser \cite{gao2022unsupervised} & $\checkmark$ & 38.2  & 87.3  & 98.4  & \textbf{75.8}  & 84.3  & 89.0  & 76.8  & 83.2  & 87.3   \\ 
    SimLM \cite{wang2022simlm} & $\checkmark$ & 39.1  & 87.3  & 98.6  & 73.6  & 84.3  & 89.0  & 78.1  & 83.8  & 87.6   \\ 
    RetroMAE \cite{liu2022retromae} & $\checkmark$ & 39.3  & 87.0  & 98.5  & 74.4  & 84.4  & 89.4  & 78.9  & 84.5  & \textbf{88.0}   \\ 
    \midrule
    \textbf{BoW Prediction} & ~ & \textbf{40.1}\textsuperscript{*}  & \textbf{88.7}\textsuperscript{*}  & \textbf{98.9}  & 75.3  & \textbf{84.6}  & \textbf{90.4}\textsuperscript{*}  & \textbf{79.4}\textsuperscript{*}  & \textbf{84.9}\textsuperscript{*}  & \textbf{88.0}  \\ 
\bottomrule
\end{tabular}
% }
\caption{
Results of Bag-of-Word (BoW) Prediction on MS-MARCO passage ranking, Natural Question, and Trivia QA tasks. The best scores are marked in bold. \textsuperscript{*} means significant improvement over the best results of previous works. ($p \leq 0.05$)
}
\label{table_results_main}
\end{table*}

\begin{table*}[!ht]
\centering
\resizebox{\linewidth}{!}{
\begin{tabular}{l|c|cc|cc|cc}
\toprule
    ~ & ~ & \multicolumn{2}{c|}{\textbf{Data Process}} & \multicolumn{2}{c|}{\textbf{Additional Decoder}} & \multicolumn{2}{c}{\textbf{Training Speed}}  \\ 
    \textbf{Pre-train Task} & \textbf{Architecture} & \textbf{Complexity} & \textbf{Time(s)} $\downarrow$ & \textbf{Complexity} & \textbf{GPU Time(s)} $\downarrow$ & \textbf{Sample per second} $\uparrow$ & \textbf{Degeneration} $\downarrow$ \\ 
    \midrule
    Further Pre-train & Encoder & $O(n)$ & 0.0476  & - & - & 269.708 & -  \\ 
    \midrule
    Auto-Encoding \cite{gao2021condenser, gao2022unsupervised, liu2022retromae, wu2023contextual} & Encoder-Decoder & $O(n)$ & 0.0940  & $O(n^2)$ & 0.0013  & 222.658 & 17.4\%  \\ 
    Auto-Regression \cite{lu2021less} & Encoder-Decoder & $O(n)$ & 0.0636  & $O(n^2)$ & 0.0030  & 215.136 & 20.2\%  \\ 
    Enhanced Decoding \cite{liu2022retromae} & Encoder-Decoder & $O(n^2)$ & 5.6261  & $O(n^2)$ & 0.0012  & 85.797 & 68.2\%  \\ 
    \midrule
    \textbf{BoW Prediction} & Encoder & $O(n)$ & \textbf{0.0533}  & - & \textbf{0.0002}  & \textbf{266.359} & \textbf{1.2\%} \\ 
\bottomrule
\end{tabular}
}
\caption{
Comparisons of architecture, data process, additional decoder, and training speed of different pre-training methods. All measurements are conducted on the same environment with a single NVIDIA H800 GPU. Note that Bag-of-Word (BoW) prediction does not involve additional decoders, thus we report the GPU computation time of the prediction task. The original implementation of enhanced decoding \cite{liu2022retromae} randomly generates 256 attention masks for each input text, and picks one as the final mask used in the pre-training. And thus it has $O(n^2)$ data process complexity.
}
\label{table_efficency}
\end{table*}

\paragraph{\textbf{Token Compositions of Dense Representation}}
MAE-style pre-training optimizes the sentence representations through bottleneck-connected encoder-decoder architecture. Although dense representations are typically highly feature-condensed and hard to interpret as human-readable format, inspired by \cite{ori2023dpr_vocab_distribution}, we can still try to visualize the dominating tokens contained in dense representations by projecting with original BERT language model heads (LM-head). To inspect the changes in sentence representations after MAE-style pre-training, we project the representations at the \texttt{[CLS]} token positions $h_{last}^{CLS}$ to the vocabulary space $h_{proj} \in vocab\_size$ and analyze the composition of top-$k$ tokens. 

\begin{equation}
    h_{proj} = h_{last}^{CLS} \cdot E^T
\end{equation}

Specifically, as is demonstrated in Algorithm \ref{algorithm_input_token_coverage}, we first take the top-$k$ dominating tokens from $h_{proj}$ and calculate the ratio of the above tokens appearing in the tokenized input token set $T = \{ t_0, t_1, .., t_n \}$. This ratio is called the input token coverage ratio. The statistics of the above pre-training checkpoints are plotted in Figure \ref{token_logits_01}.
Compared to the original BERT checkpoint, auto-encoding and further pre-training do not contribute to the input token coverage ratio. However, MAE-style pre-training with enhanced decoding significantly enhances the dominating token coverage of input texts. Our interpretation pipeline and an example of all pre-training schemas are displayed in Figure \ref{logits_view}. Here we can also observe higher input token coverage of pre-training with enhanced decoding, i.e. more hits ✓ on the inputs, compared to all other pre-training methods. Considering the highest retrieval performances of enhanced decoding, which are shown in Table \ref{table_mae_eval_loss_msmarco} and \ref{table_mae_eval_loss_dpr}, we hypothesize that the increased retrieval abilities may come from enhanced dominating token coverage of input texts.

% \begin{figure}[t!]
% \centering
% \includegraphics[width=7cm]{figs/finetune_pipeline.pdf}
% \caption{
% Supervised fine-tuning pipeline of our method. The retrievers (s1, s2) are fine-tuned by the ANCE \cite{xiong2020ance} fashion.
% }
% \label{finetune_pipeline}
% \end{figure}

\subsection{Pre-training with Bag-of-Word Prediction}
Building upon the above observations about token compositions and domain adaptation abilities, here we propose a new pre-training schema that explicitly removes the mandatory needed decoder in existing MAE-style pre-training. As is shown in Algorithm \ref{algorithm_bag_of_word}, the sentence representations at the \texttt{[CLS]} token position are first projected to the vocabulary space with the BERT language-model heads (i.e., its embedding matrix $E^T$). Then we apply a multi-label cross-entropy loss to predict the input token set $T$ directly. Bag-of-Word means the label $T$ is a set of tokenized input text. For simplicity, it is not related to positional information, similar to the traits of sparse retrieval \cite{robertson2009probabilistic}. % TODO 所以这里只是预测所有的token，不是真正的bag of word？如果是这样，最好换个词，bag of word已经是一个很重要的team了。

\begin{equation}
\label{euqation_bow}
\begin{aligned}
    \mathcal{L}_{Dec} = CE(h_{last}^{CLS} E^T, T)
\end{aligned}
\end{equation}

Then the final objective function of our method comprises both the encoder MLM loss and the Bag-of-Word prediction loss.

\begin{equation}
\begin{aligned}
    \mathcal{L} = \mathcal{L}_{Enc} + \mathcal{L}_{Dec}
\end{aligned}
\end{equation}

\begin{table*}[!ht]
\centering
% \resizebox{\linewidth}{!}{
\begin{tabular}{l|ccc|ccc|ccc}
\toprule
    ~ & \multicolumn{3}{c|}{\textbf{MS-MARCO}} & \multicolumn{3}{c|}{\textbf{Natural Question}} & \multicolumn{3}{c}{\textbf{Trivia QA}}  \\ 
    \textbf{Model} & MRR@10 & R@50 & R@1k & R@5 & R@20 & R@100 & R@5 & R@20 & R@100  \\ 
    \toprule
    \multicolumn{10}{l}{\textbf{Fine-tuned with BM25 Negatives}} \\
    \midrule
    BERT & 33.7  & 80.9  & 96.4  & 68.2  & 79.8  & 87.0  & 70.9  & 80.2  & 85.7   \\
    Further Pre-train & 37.2  & 86.0  & 98.2  & 71.4  & 81.9  & 88.2  & 75.1  & 82.3  & 86.8   \\
    \midrule
    Auto-Encoding & 37.6  & 86.3  & 98.5  & 71.6  & 82.5  & 88.8  & 76.4  & 83.2  & 87.2   \\
    Auto-Regression & 38.0  & 86.9  & \textbf{98.7}  & 71.4  & 82.8  & 88.8  & 76.7  & 83.6  & 87.4   \\
    Enhanced Decoding & 37.0  & 86.7  & 98.4  & \textbf{72.3}  & \textbf{83.0}  & 88.9  & 77.2  & 83.5  & 87.4   \\
    \midrule
    \textbf{BoW Prediction} & \textbf{38.4}\textsuperscript{*}  & \textbf{87.5}\textsuperscript{*}  & \textbf{98.7}  & \textbf{72.3}  & 82.9  & \textbf{89.3}\textsuperscript{*}  & \textbf{77.7}\textsuperscript{*}  & \textbf{83.9}\textsuperscript{*}  & \textbf{87.6}  \\
    \toprule
    \toprule
    \multicolumn{10}{l}{\textbf{Fine-tuned with Mined Negatives}} \\
    \midrule
    BERT & 36.1  & 83.5  & 96.9  & 72.5  & 83.1  & 88.8  & 75.0  & 82.2  & 86.8   \\
    Further Pre-train & 39.2  & 87.7  & 98.5  & 74.0  & 84.0  & 89.7  & 77.4  & 83.8  & 87.6   \\
    \midrule
    Auto-Encoding & 39.3  & 87.9  & 98.7  & 74.9  & 84.5  & 90.0  & 77.5  & 84.0  & 87.6   \\
    Auto-Regression & 39.7  & 88.3  & \textbf{98.9}  & 74.8  & 84.4  & 90.1  & 78.8  & 84.6  & 87.9   \\
    Enhanced Decoding & 39.3  & 87.0  & 98.6  & 74.4  & 84.5  & 90.3  & 79.0  & 84.5  & \textbf{88.1}   \\
    \midrule
    \textbf{BoW Prediction} & \textbf{40.1}\textsuperscript{*}  & \textbf{88.7}\textsuperscript{*}  & \textbf{98.9}  & \textbf{75.3}\textsuperscript{*}  & \textbf{84.6}  & \textbf{90.4}  & \textbf{79.4}\textsuperscript{*}  & \textbf{84.9}  & 88.0  \\
\bottomrule
\end{tabular}
% }
\caption{
Retrieval performances of different pre-training schema with BM25 negatives and mined negatives. \textsuperscript{*} means significant improvement over the best results of reproduced MAE pre-training baselines. ($p \leq 0.05$)
}
\label{table_bm25_mine_negs}
\end{table*}

There are several benefits of the design behind our method. First of all, Bag-of-Word prediction is a direct compression of input lexicon information into dense representations with extreme simplicity, which merely utilizes the existing BERT language model head for projection. Secondly, it removes the need for extra Transformers-based decoder layers, data preprocessing, or extensive data masking, which introduces \textbf{ZERO} extra computational complexity, compared to other expensive MAE-style pre-training. Moreover, different from previous MAE-style pre-training modifications, which mostly focus on extensive modifications to the decoder tasks \cite{liu2022retromae, wang2022simlm, wu2023contextual, zhou2022master}, Bag-of-Word pre-training only involves a simple lexicon prediction task, which bridges the condensed representations and lexicon traits. Such a pre-training schema is highly interpretable and easy to implement.

\section{Experiments}
This section introduces the experiment settings and performances of Bag-of-Word prediction. 
% used in Bag-of-Word prediction pre-training. Then we analyze the retrieval performances over several large-scale retrieval benchmarks and compare the pre-training efficiencies among various pre-training objects. % TODO 全文的表格和figure重新排版一下，确保每个文字对应的table或者figure尽量都在一页里头，好找。

\subsection{Experiment Settings}
\paragraph{\textbf{Pre-training}}
Wikipedia \cite{wikidump} and Bookcorpus \cite{zhu2015bookcorpus}, the original BERT training sets, are used as pre-training corpora, which are commonly adopted in previous retrieval pre-trainings \cite{gao2021condenser, liu2022retromae, lu2021less}. The \texttt{BERT-base-uncased} checkpoint is used to initialize the encoder for saving computation. The model is pre-trained using the AdamW optimizer for 20 epochs, with a learning rate of 3e-4, a batch size of 2048, and a maximum sequence length of 512. Following previous in-domain pre-training settings \cite{gao2021condenser, liu2022retromae, lu2021less, wang2022simlm, yu2024corpus_specific_vocabularies}, MS-MARCO corpus \cite{tri2016msmarco} is also used when evaluating the MS-MARCO benchmark for making fair comparisons with existing works \cite{gao2021condenser, liu2022retromae, wang2022simlm}. All pre-training hyper-parameters are kept the same, except the maximum sequence length is cut to 144. 

\paragraph{\textbf{Evaluation}}
We conduct supervised evaluations on several large-scale web search and open-domain QA datasets, including MS-MARCO passage ranking \cite{tri2016msmarco}, Natural Questions (NQ) \cite{tom2019nq} and TriviaQA \cite{Mandar2017TriviaQA}. 
% MS-MARCO \cite{tri2016msmarco} is one of the large-scale search datasets consisting of real Bing questions and human-annotated answers with 8.8 million passages. Natural Questions (NQ) is a large open-domain question-answering dataset containing 320K examples from Google search and corresponding Wikipedia pages, and TriviaQA is a reading comprehension dataset containing over 650K question-answer-evidence triples. 
The metrics of Mean Reciprocal Rank (MRR@k) and Recall (R@k) are reported by following \cite{gao2022unsupervised}. 
We fine-tune MS-MARCO with a batch size of 64, a learning rate of 2e-5, and a negative size of 15 for 3 epochs. Following \cite{karpukhin2020dpr}, we utilize the DPR-Wiki versions of NQ and TriviaQA benchmarks and evaluate retrieval performances. We fine-tuned NQ and TriviaQA with the same hyper-parameters, except that the learning rate is 5e-6, and the number of training epochs is 20. 
% As is shown in Figure \ref{finetune_pipeline}, the two-stage 
Retrievers are fine-tuned with ANCE manners \cite{xiong2020ance} by following \cite{gao2021condenser, gao2022unsupervised, liu2022retromae}.
Following \cite{gao2021condenser, liu2022retromae}, both BM25 negatives and retriever-mined negatives from previous stage are used in fine-tuning stages for fair comparisons. A large-scale heterogeneous zero-shot benchmark BEIR \cite{thakur2021beir} is also used to evaluate the out-of-domain transferability of our method. Normalized Discounted Cumulative Gain (NDCG@10) is used as the metric for BEIR.
% All datasets are evaluated on the test sets, except that MS-MARCO is on the development set since its test set is not publicly available.

\paragraph{\textbf{Baselines}}
We include multiple public sparse and dense retrieval baselines for comparisons. Sparse retrieval baselines include BM25 \cite{robertson2009probabilistic} and SPLADE++ \cite{Thibault2022splade++}. Dense retrieval baselines are composed of multiple well-known pre-training designs, including MAE-style pre-training SEED \cite{lu2021less}, Condenser \cite{gao2021condenser}, coCondenser \cite{gao2022unsupervised}, SimLM \cite{wang2022simlm}, RetroMAE \cite{liu2022retromae}, and other dense retrieval modifications, like DPR \cite{karpukhin2020dpr}, ANCE \cite{xiong2020ance}, RocketQA \cite{qu-etal-2021-rocketqa}. Distillations from re-rankers are not included as their retrieval performances rely more on the capability of re-ranker teachers, while in our study we focus on inspecting the dense retrieval per-training of retrievers. Particularly, we reproduce multiple MAE-style pre-training methods with same hyperparameters and compare with their performances and training efficiencies, including Auto-Encoding \cite{liu2022retromae}, Auto-Regression \cite{lu2021less}, and Enhanced Decoding \cite{liu2022retromae}. 

Other data augmentation techniques, such as context-based \cite{wu2023contextual}, expansion-based \cite{wu2022query-as-context}, mutual information-based \cite{zehan2023cdmae} or multi-task \cite{zhou2022master} pre-training, are not included here for fair comparisons. Our proposed Bag-of-Word prediction is orthogonal to them, and we will leave the combinations of our method with various augmentation techniques to further work. Hybrid retrievals \cite{liu2023retromaev2, wu2023cotmaev2} are not included in the baselines, as our work only focuses on the pre-training for dense retrieval.

\begin{figure}[t!]
\centering
\includegraphics[width=\linewidth]{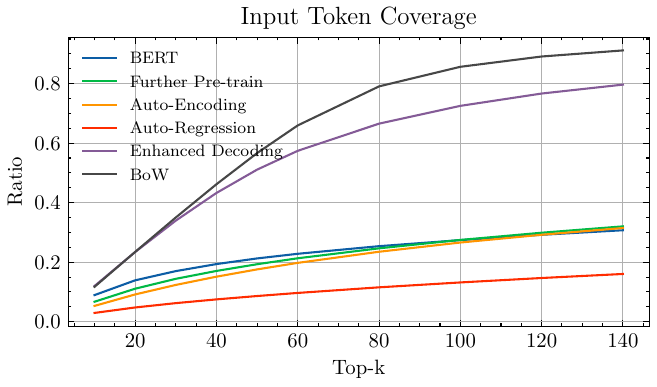}
\caption{
Input token coverage about Top-k tokens of dense representation after Bag-of-Word prediction pre-training.
}
\label{token_logits_02}
\end{figure}

\begin{figure}[t!]
\centering
\includegraphics[width=\linewidth]{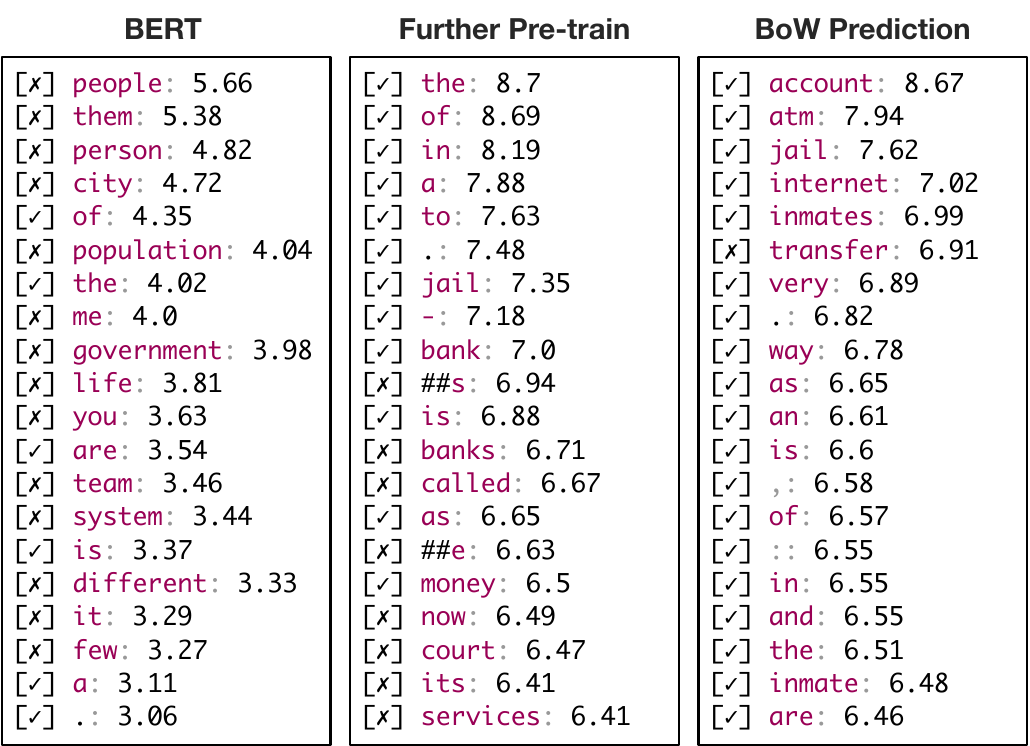}
\caption{
Examples of the compositions of Top-20 tokens of dense representation after pre-training with Bag-of-Word prediction.
}
\label{logits_view_bow}
\end{figure}

\subsection{Main Results}
\paragraph{\textbf{Retrieval Performances}}
Pre-training with Bag-of-Word prediction achieves state-of-the-art retrieval performances on several large-scale retrieval benchmarks. As is displayed in Table \ref{table_results_main}, our method outperforms the strongest baseline by +0.8\% (MRR@10) and +1.7\% (R@50) on MS-MARCO, +1.0\% (R@100) on NQ and +0.5\% (R@5 and R@20) on TriviaQA. Note that compared with the existing MAE-style pre-training, our method explicitly removes the mandatory need for the additional decoder and outperforms several MAE retrieval baselines. The performance improvements completely come from the incorporation of the Bag-of-Word pre-training task, other than data augmentation or task ensembles. Our method works with extreme simplicity and is robust to various retrieval domains.

\paragraph{\textbf{Pre-training Efficiency}}
Bag-of-Word prediction pre-trains dense retrieval models with higher speed and fewer CPU/GPU complexities. We compare the training efficiencies with a single NVIDIA H800 GPU. As is shown in Table \ref{table_efficency}, compared with MAE-based encoder-decoder structures, our method removes the need for the additional decoder, which saves > 83\% GPU computation time on the decoding side. Additionally, our method does not involve extensive masking strategies, which saves 99\% CPU time for the data process compared to enhanced decoding. 
Overall, the Bag-of-Word prediction has a very slight training speed degeneration and achieves 67\%, 19\%, and 16.2\% training speedup compared with enhanced decoding, auto-regression, and auto-encoding.

\begin{table}[!t]
\centering
\resizebox{\linewidth}{!}{
\begin{tabular}{l|c|ccc|c}
\toprule
    ~ & ~ & \multicolumn{3}{c|}{\textbf{Encoder Mask Ratios}} & ~ \\ 
    \textbf{Datasets / nDCG@10} & Baseline & 15\% & 30\% & 45\% & wo/ BoW \\
    \toprule
    TREC-COVID & 65.69  & 68.82  & 59.40  & 64.60  & 67.01  \\
    BioASQ & 27.51  & 39.44  & 40.06  & 39.43  & 34.99  \\
    NFCorpus & 27.41  & 31.19  & 31.45  & 30.85  & 29.41  \\
    NQ & 46.75  & 50.52  & 48.68  & 48.18  & 47.59  \\
    HotpotQA & 48.28  & 64.10  & 63.64  & 62.47  & 55.25  \\
    FiQA-2018 & 24.43  & 30.87  & 30.77  & 30.48  & 28.31  \\
    Signal-1M & 18.86  & 28.07  & 29.33  & 28.97  & 26.71  \\
    TREC-NEWS & 36.61  & 41.10  & 41.17  & 40.20  & 40.01  \\
    Robust04 & 37.51  & 44.14  & 44.46  & 43.67  & 42.04  \\
    ArguAna & 33.33  & 41.93  & 44.94  & 44.29  & 49.63  \\
    Touche-2020 & 25.55  & 21.22  & 17.18  & 18.39  & 20.59  \\
    CQADupStack & 27.28  & 32.74  & 32.96  & 32.59  & 31.38  \\
    Quota & 78.43  & 85.85  & 85.75  & 85.84  & 86.29  \\
    DBPedia & 31.95  & 38.72  & 38.27  & 38.84  & 36.27  \\
    SCIDOCS & 12.14  & 15.92  & 15.70  & 15.55  & 14.36  \\
    FEVER & 69.86  & 74.61  & 72.27  & 72.75  & 70.72  \\
    Climate-FEVER & 20.56  & 24.01  & 25.78  & 25.56  & 24.39  \\
    SciFact & 50.66  & 64.63  & 65.52  & 64.84  & 58.81  \\
    \midrule
    \textbf{Average} & 37.93  & \textbf{44.33}\textsuperscript{*}  & 43.74\textsuperscript{*}  & 43.75\textsuperscript{*}  & 42.43  \\
    \textbf{Relative Gains} & - & \textbf{+6.39}  & +5.81  & +5.81  & +4.50 \\
\bottomrule
\end{tabular}
}
\caption{
Ablation study on the heterogeneous BEIR benchmark for our Bag-of-Word prediction pre-training method, with various encoder mask ratios or removing the proposed Bag-of-Word prediction task. \textsuperscript{*} means significant improvements over the wo/BoW settting. ($p \leq 0.05$)
}
\label{table_beir}
\end{table}

\section{Ablation Studies}
\subsection{Performances with Different Negatives}
We compare the retrieval performances of different pre-training schemas and report the results with different negative training data in Table \ref{table_bm25_mine_negs}. The retrievers are trained with BM25 negatives and retriever-mined negatives, respectively, by following previous supervised fine-tuning pipelines \cite{xiong2020ance, gao2021condenser}. It can be seen that even with different negatives, Bag-of-Word prediction pre-training surpasses all MAE-style pre-training, MLM further pre-training, and BERT baselines, under the same pre-training and fine-tuning pipelines. This further verifies that our pre-training method is simple but more effective than other MAE pre-training schemas.

\subsection{Input Token Coverage}
As is presented in Section \ref{method_interprete}, we visualize the input token coverage about top-k tokens by projecting dense representations to the vocabulary space via the BERT language model heads. As is shown in Figure \ref{token_logits_02}, Bag-of-Word has a higher input token coverage ratio than other pre-training methods. It implies that Bag-of-Word prediction has higher compression efficiency of lexicon information than MAE-style pre-training with its direct prediction task. The higher retrieval performances are also attributed to the direct compression of the input token into dense representations. Examples in Figure \ref{logits_view_bow} further verify that Bag-of-Word prediction enables dense representation to give higher frequencies to input tokens, especially representative tokens, thus achieving higher retrieval performances.

\subsection{Transferability and Encoder Mask Ratios}
Previous study \cite{liu2022retromae, wu2023contextual} reported that properly increasing the mask ratio for the encoder can lead to higher retrieval performances. Hence we perform pre-training with Bag-of-Word prediction under different encoder mask ratios and report the performances on the heterogeneous BEIR benchmarks. BEIR benchmarks include 18 out-of-domain retrieval datasets across different domains, sizes, and topics, which is suitable for evaluating the transferability of retriever models. As is shown in Table \ref{table_beir}, pre-training with Bag-of-Word prediction achieves relatively +6.39\% performance gains over the BERT baseline. On the contrary, removing the Bag-of-Word prediction task and applying further pre-training with pure MLM loss will incur 1.89\% (6.39\% $\rightarrow$ 4.50\%) performance degeneration. Here we also observe that increasing the encoder mask ratio from 15\% to 30\% or 45\% does not yield higher retrieval performances.

\section{Conclusion}
Masked auto-encoder pre-training has become a prevalent technique for initializing and enhancing dense retrieval systems. However, the common utilization of additional Transformers-based decoders incurs significant computational costs. And the underlying reasons for its effectiveness are not well explored. In this paper, we first reveal that MAE pre-training with enhanced decoding significantly increases the input lexicon coverage, via visualizing the dense representations. Then we propose a direct modification to traditional MAE by replacing the decoder with merely a Bag-of-Word prediction task. The Bag-of-Word prediction task is extremely simple but effective for the pre-training of dense retrieval systems. Our method achieves state-of-the-art retrieval performances without requiring any additional Transformers decoders, data preprocessing, or extensive data masking strategies. It is also a direct compression of lexicon information to dense representations, which is highly interpretable and easy to implement.

\section*{Limitations}
In this study, we explore the direct application of Bag-of-Word prediction in dense retrieval pre-training. There are also effective modifications about data augmentation or task-ensembling techniques, such as context sampling \cite{wu2023contextual}, document expansion \cite{Rodrigo2019doc2query, wu2023contextual}, mutual information masking \cite{zehan2023cdmae} and multi-task \cite{zhou2022master}. Bag-of-Word prediction focuses on pre-training paradigm innovation, which is orthogonal to them. Combinations of Bag-of-Word prediction with those data augmentation or task-ensembling techniques will be likely to further boost the downstream retrieval performances, and we leave this to our future work.

\bibliographystyle{ACM-Reference-Format}
\bibliography{sample-base}

% \appendix

\end{document}